\documentstyle[pre,aps,twocolumn,epsf]{revtex}
\tighten

\begin{document} 
\draft

\title{
Phase transitions in the Potts spin glass model
}  
\author{
Giancarlo Franzese$^{1,2}$ and Antonio Coniglio$^{1,2,3}$
}
\address{
$^1$ Dipartimento di Scienze Fisiche,
Universit\`a di Napoli, Mostra d'Oltremare Pad.19 I-80125 Napoli
Italy\\
$^2$ INFM - unit\`a di Napoli ~~~ 
$^3$ INFN - sezione di Napoli
} 

\date{\today}

\maketitle 

\begin{abstract} 
We have studied the Potts spin glass with 2-state Ising spins and 
$s$-state Potts variables using a cluster Monte Carlo 
dynamics. The model recovers the $\pm J$ Ising spin glass (SG) 
for $s=1$ and exhibits for all $s$ a SG transition at $T_{SG}(s)$ and
a percolation transition at higher temperature $T_p(s)$.
We have shown that for all values of 
$s\neq 1$ at $T_p(s)$ there is a thermodynamical transition in the universality 
class of a ferromagnetic $s$-state Potts model.
The efficiency of the cluster dynamics is compared with that of standard 
spin flip dynamics.
\end{abstract} 
\pacs{PACS numbers: 75.10.Nr, 64.60.Ak, 02.70.Lq }



\section{Introduction}

In nature there are many examples of glassy systems,
{\it i.e.} complex systems that exhibit a
very slow dynamics that prevents them to reach the equilibrium state.
In this class there is a large variety of systems, like   
real glasses, spin glasses, supercooled liquids, polymers, 
granular material, colloids, ionic conductors, orientational glasses,
vortex glasses \cite{Sitges}.
A common feature of these systems is the {\em frustration}, that is a 
competition due to geometry or energy constrains.

Experiments on glasses \cite{Angell} show that in a temperature 
driven transition precursor phenomena starts at a temperature well above 
the ideal glass transition. This temperature 
depends on experimental conditions and is the onset of some dynamical 
anomalies, like non-exponential relaxation functions, anomalous 
diffusion, cooling rate dependent density.  
Analogous phenomena are present 
in spin glasses (SG) \cite{sg}, that are  a generalization of Ising model 
where random distributed ferromagnetic and antiferromagnetic interactions give 
rise to frustration. In particular, experiments \cite{Mezei} 
and numerical simulations \cite{Ogielski} have shown
non-exponential autocorrelation functions below a
temperature $T^*$ well above the transition temperature $T_{SG}$, while
above $T^*$ only exponential relaxation functions are seen.

To explain this phenomenon, Randeria {\it et al.} \cite{Randeria} have 
suggested that in the SG the
non-exponential regime starts at the Griffiths temperature $T_c$, {\it 
i.e.} the critical temperature of the ferromagnetic model, due to the 
presence of randomly large unfrustrated regions, allowed by the 
disorder in the SG interactions. 
Campbell {\it et al.} \cite{Campbell} have proposed an alternative hypothesis 
in which the onset $T^*$ of non-exponential behavior coincides with
the percolation temperature $T_p$ of Fortuin-Kasteleyn 
\cite{FK} Coniglio-Klein \cite{CK} (FK-CK) clusters (defined below). 
The idea of Ref.\cite{Campbell} is that in the SG the accessible 
phase space is simply connected
above $T_p$, while it is not below $T_p$. Therefore for $T<T_p$ 
a local Monte Carlo (MC) dynamics performs a random 
walk on a ramified percolating-like structure with many time scales, 
giving rise to non-exponential relaxation functions. 

Since $T_p$ is less than, but close to, $T_c$ and $T^*$ is difficult to localize, 
it is not possible to exclude the $T^*=T_c$ hypothesis.
However numerical results \cite{Fierro,FF} on fully frustrated models, without disorder,
where the Randeria {\it et al.} argument does not apply, are consistent with the 
$T^*=T_p$ scenario. 
In Refs.\cite{Fierro,FF} this result is given both for spin-flip and
bond-flip dynamics. The latter is strictly related to the bond {\em frustrated
percolation} (FP) where {\it frustrated loops}, {\it i.e.} 
closed paths of connected bonds covering an odd number of negative interactions,
have zero weight \cite{Fierro}.
The FP model and the $\pm J$ SG model can be recovered as particular cases of   
a $2s$-state Potts spin glass (PSG) \cite{CdLMP} respectively for $s=1/2$ and
$s=1$. 

In this paper we will show numerical results on the static properties 
of the PSG model in 2D. 
For any $s$ the model exhibits a SG transition at a temperature $T_{SG}(s)$ (in 
2D $T_{SG}=0$ for any $s$) and a FK-CK percolation transition at higher 
temperature $T_p(s)$. 
For $s\neq 1$ the higher transition corresponds to a real thermodynamical
transition of a $s$-state Potts model \cite{Wu}. 
To make the algorithm faster 
we use the Swendsen and Wang (SW) \cite{SW} 
cluster MC dynamics that prevent the slowing down for temperatures 
near $T_p(s)$.
Dynamical properties of the model
for $s=2$ are given in Ref.\cite{breve}, where we have shown that 
autocorrelation times diverge at $T_p$ and, as in SG model, non-exponential
relaxation functions are present below $T_p$. 

These results on SG model \cite{Campbell}, FP model \cite{Fierro}, fully 
frustrated systems \cite{FF} and PSG model \cite{breve} suggest
that the FK-CK percolation may play a role in the context of precursor phenomena,
since below $T_p$ the frustration is present at all length scales by means
of FK-CK percolating cluster, that cannot include frustrated loops.

In Sec. II we define the Hamiltonian 
and review some theoretical results. 
In Sec. III we present the numerical results for the model in finite dimensions.
In Sec. IV we define the SW dynamics and compere it with the 
spin flip dynamics, verifying the efficiency of the cluster dynamics.
In Sec. V we give the conclusions.

\section{Hamiltonian Formalism}

The Potts spin glass (PSG) model is defined by the Hamiltonian 
\begin{equation}
H=-sJ\sum_{\langle i,j \rangle} [\delta_{\sigma_i \sigma_j}
(\epsilon_{i,j}S_i S_j+1)-2]
\label{hamiltonian}
\end{equation}
where to each lattice site is associated an Ising spin $S_i=\pm 1$ and a 
$s$-state Potts spin $\sigma_i=1, \dots , s$. 
The sum is extended over all NN sites,   
$\epsilon_{i,j}=\pm 1$ is a random quenched variable and
$J$ is the strength of interaction.
The model is a superposition of a ferromagnetic $s$-state Potts model 
\cite{Wu} and a $\pm J$ Ising SG \cite{sg} and 
for $\delta_{\sigma_i \sigma_j}=1$ ({\it i.e.} $s=1$) recovers the $\pm J$ 
Ising SG Hamiltonian.

Following Ref.\cite{CdLMP} it is possible to define FK-CK clusters on this
model, activating a bond, between NN sites with both SG interaction and Potts 
interaction satisfied, with probability 
\begin{equation}
p=1-e^{-2sJ/k_BT} ~,
\label{p}
\end{equation} 
and defining a cluster as the maximal set of connected bonds.

For a given set of interaction $\{\epsilon_{i,j}\}$ it 
is possible to shown \cite{CdLMP} that $Z$ can 
be expressed in terms of bond configurations $C$
\begin{equation}
Z\{\epsilon_{i,j}\}=\sum_{\{S_i,\sigma_i\}}e^{-H/k_BT}=\sum_C W_s(C)
\label{partition_function}
\end{equation}
where $W_s(C)=0$ if $C$ includes any frustrated loop, otherwise
\begin{equation}
W_s(C)=p^{|C|}(1-p)^{|A|}(2s)^{N(C)}
\label{W_q}
\end{equation}
where $p$ is given in eq. (\ref{p}), $N(C)$ is the number of clusters in 
the configuration $C$, $|C|$ is the number of bonds and $|C|+|A|$ is total 
number of interactions. Let us observe that, while the Hamiltonian 
(\ref{hamiltonian}) is defined only for integer values of $s$,  the eq.
(\ref{partition_function}) is meaningful for every values of $s$ and that 
for $s=1/2$ the eq. (\ref{partition_function}) gives the partition 
function of bond FP where a bond configuration without frustrated loops 
has a weight $W(C)=e^{\beta\mu |C|}$ with $\beta\mu=\ln(e^{\beta J}-1)$,
while a bond configuration with frustrated loops has a zero weight.
Furthermore in the limit $s\rightarrow 0$ eq. (\ref{partition_function}) 
gives the partition function of the {\em tree percolation } \cite{Wu}
where any bond configuration with a loop is excluded.

Exact renormalization group (RG) analysis on hierarchical lattice 
\cite{Pezzella} has predicted for the PSG 
two critical temperatures. The lower 
temperature $T_{SG}(s)$ corresponds to a SG transition in the universality class 
of $\pm J$ Ising SG, and the higher $T_p(s)$ to a 
percolation transition in the universality class of a ferromagnetic $s$-state 
Potts model.
Same results are given for the fully frustrated version of the model studied 
with a mean field approach \cite{dLP}. 

Looking at the partition function (\ref{partition_function}) one should expect
a singularity at $T_p(s)$, for any $s\neq 0$, due to the 
singularity in the number of cluster $N(C)$.
Nevertheless this singularity has never been observed in the case of SG 
($s=1$). In fact, the RG calculations in the case $s=1$ show a singularity 
at $T_{SG}$ for the SG free energy and no singularity at $T_p(1)$. 
This result is interpreted in Ref.\cite{Pezzella} supposing that the free energy of 
the Hamiltonian (\ref{hamiltonian}) has the form 
\begin{equation}
F_s(T)-F_s(T_p)\sim A(s)(T-T_p(s))^{2-\alpha(s)}
\label{free_energy}
\end{equation}
where $A(s)$ is an amplitude which vanishes for $s\rightarrow 1$ and 
$\alpha(s)$ is the specific heat exponent.

\section{Monte Carlo results}

We have done our simulation using the Swendsen and Wang
(SW) Monte Carlo cluster dynamics \cite{SW} described in Sec. IV.
As we will show, the SW dynamics is 
faster then standard local dynamics, like spin flip dynamics, but suffers of 
slowing down near the SG critical temperature.  
Nevertheless, since the SG transition in 2D occurs at $T_{SG}=0$ 
\cite{BathY} and we are interested in study the system near the percolation
transition at $T_p>T_{SG}$, the SW dynamics is particularly indicated.

We have performed numerical simulation of the PSG model for $s=2,7,50$ on a 
2D square lattice with linear sizes ranging from $L=10$ to $60$ lattice steps 
and with quenched 
random interaction configurations $\{\epsilon_{ij}\}$.
Defining as MC step an update of all the spins
of the system, we have discarded the data of the first 7~500 MC steps and have
collected data over 15~000, 25~000 or 50~000 MC steps, depending on the
temperatures and sizes.

For each $s$ we have calculated the Binder parameter for the energy density
$E$ \cite{Binder} defined as
\begin{equation}
V=1-\frac{\langle E^4 \rangle}{3\langle E^2 \rangle^2}
\label{V}
\end{equation}
where the symbol $\langle \cdot \rangle$ stands for the thermal average. This
quantity allows to localize the transition and to
distinguish between first order and second
order phase transition. In fact, for a second order phase transition in the
limit $L\rightarrow \infty$ it is $V=2/3$ for all temperatures, while for
a first order phase transition it is
\begin{equation}
\frac{2}{3}-V^{\mbox{min}}=\frac{1}{3}\frac{(E_+-E_-)^2(E_++E_-)^2}
{(E_+^2+E_-^2)^2}
\label{E+E-}
\end{equation}
where $V^{\mbox{min}}$ is the minimum value of $V$ (occurring at the phase
transition temperature) and $E_+-E_-$ is the energy jump, related to the latent
heat, at the same temperature.

To estimate the thermodynamical critical exponents for the second order
phase transition we have measured the Potts order
parameter
\begin{equation}
M=\frac{s~ \mbox{max}_i(M_i)-1}{s-1}
\label{M}
\end{equation}
(where $i=1, \dots s$, $M_i$ is the density of Potts spins in the
$i$th state), the susceptibility
\begin{equation}
\chi=\left[ \frac{\langle M^2 \rangle - \langle M \rangle^2}{N}
\right]^{1/2}
\label{chi}
\end{equation}
(where $N$ is the total number of Potts spins) and the specific heat
\begin{equation}
C_H=\left[ \frac{\langle E^2 \rangle - \langle E \rangle^2}{N}
\right]^{1/2}.
\label{cv1}
\end{equation}
Furthermore to estimate the percolation critical exponents
we have calculated the percolation probability per spin
\begin{equation}
P=1-\sum_k k n_k ~ ,
\label{P}
\end{equation}
where $k$ is the cluster size and $n_k$ is the density of clusters of size
$k$, the mean cluster size
\begin{equation}
S=\sum_k k^2 n_k ~,
\label{S}
\end{equation}
and the number of clusters
\begin{equation}
N_c=\sum_k n_k ~.
\label{Nc}
\end{equation}

\subsection{Results for $s=2$}

\begin{figure}
\begin{center}
\mbox{ \epsfxsize=8cm \epsffile{ 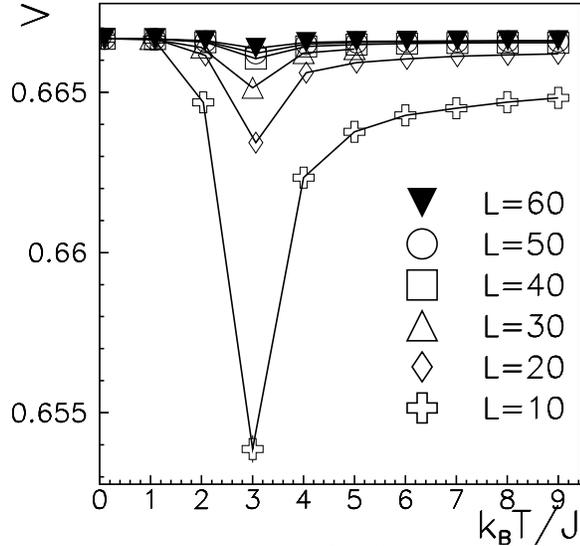 } }
\caption{PSG model for $s=2$: 
Binder parameter $V$ vs. dimensionless temperature $k_B T/J$ for $L=10\div 60$. 
Errors are smaller than the symbol sizes.  Lines are only guides for the eyes.
}
\label{V4}
\end{center}
\end{figure}

In Fig.\ref{V4} we show the Binder parameter for the case $s=2$ for system
sizes $L=10\div 60$. It is possible to see that $V$ for small sizes has a
minimum at $k_BT/J\simeq 3.0$ and that for grater sizes it becomes constant for
all temperatures, reveling a second order phase transition. Therefore we can
make standard scaling analysis \cite{Bin88} for the thermodynamical quantities.

In particular, by definition of critical exponents $\nu$ it is
\begin{equation}
\xi\sim|T-T_s|^{-\nu}
\label{nu}
\end{equation}
where $\xi$ is the correlation length and $T_s=\lim_{L\rightarrow \infty}
T_s(L)$ with $T_s(L)$ finite size critical temperature of the PSG
model. 
Analogously from the definitions of the other critical exponents $\beta$,
$\gamma$ and $\alpha$ we get
\begin{equation}
M\sim|T-T_s|^{\beta}\sim \xi^{-\beta/\nu} ~,
\label{beta}
\end{equation}
\begin{equation}
\chi\sim|T-T_s|^{-\gamma}\sim \xi^{\gamma/\nu} ~,
\label{gamma}
\end{equation}
\begin{equation}
C_H\sim|T-T_s|^{-\alpha}\sim \xi^{\alpha/\nu} ~.
\label{alpha}
\end{equation}

From standard scaling analysis applied to finite systems 
\cite{Bin88} we
expect for $M$
\begin{equation}
M\sim L^{-\beta/\nu} f_M((T-T_s)L^{1/\nu})
\label{scaling_M}
\end{equation}
where $f_M(x)$ is an universal function of the dimensionless variable $x$.
Analogous scaling functions are expected for the other thermodynamical quantities.
Tuning the values of critical exponents and $T_s$ it is possible to verify the scaling
hypothesis, as eq.(\ref{scaling_M}), from the MC data. The values for which the data
collapse give the estimates of the critical exponents and of $T_s$.

\begin{figure}
\begin{center}
\mbox{ \epsfxsize=8cm \epsffile{ 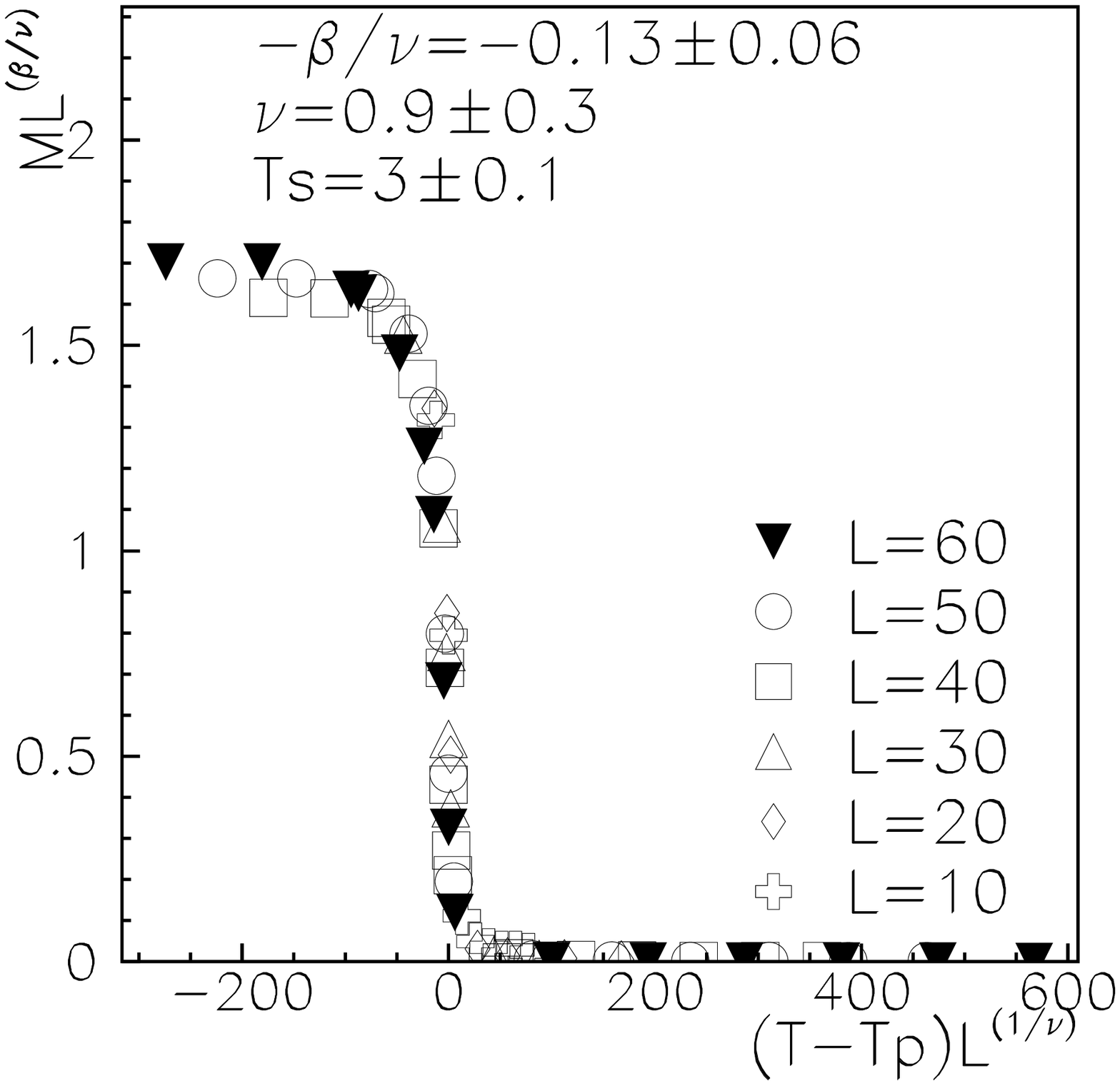 } }
\caption{PSG model with $s=2$: Data collapse for $M$ for systems sizes $L=10\div 
60$. Temperatures ate in $J/k_B$ units. The scaling parameters are given in the 
figure.
}
\label{sca_s2}
\end{center}
\end{figure}

\begin{figure}
\begin{center}
\mbox{ \epsfxsize=8cm \epsffile{ 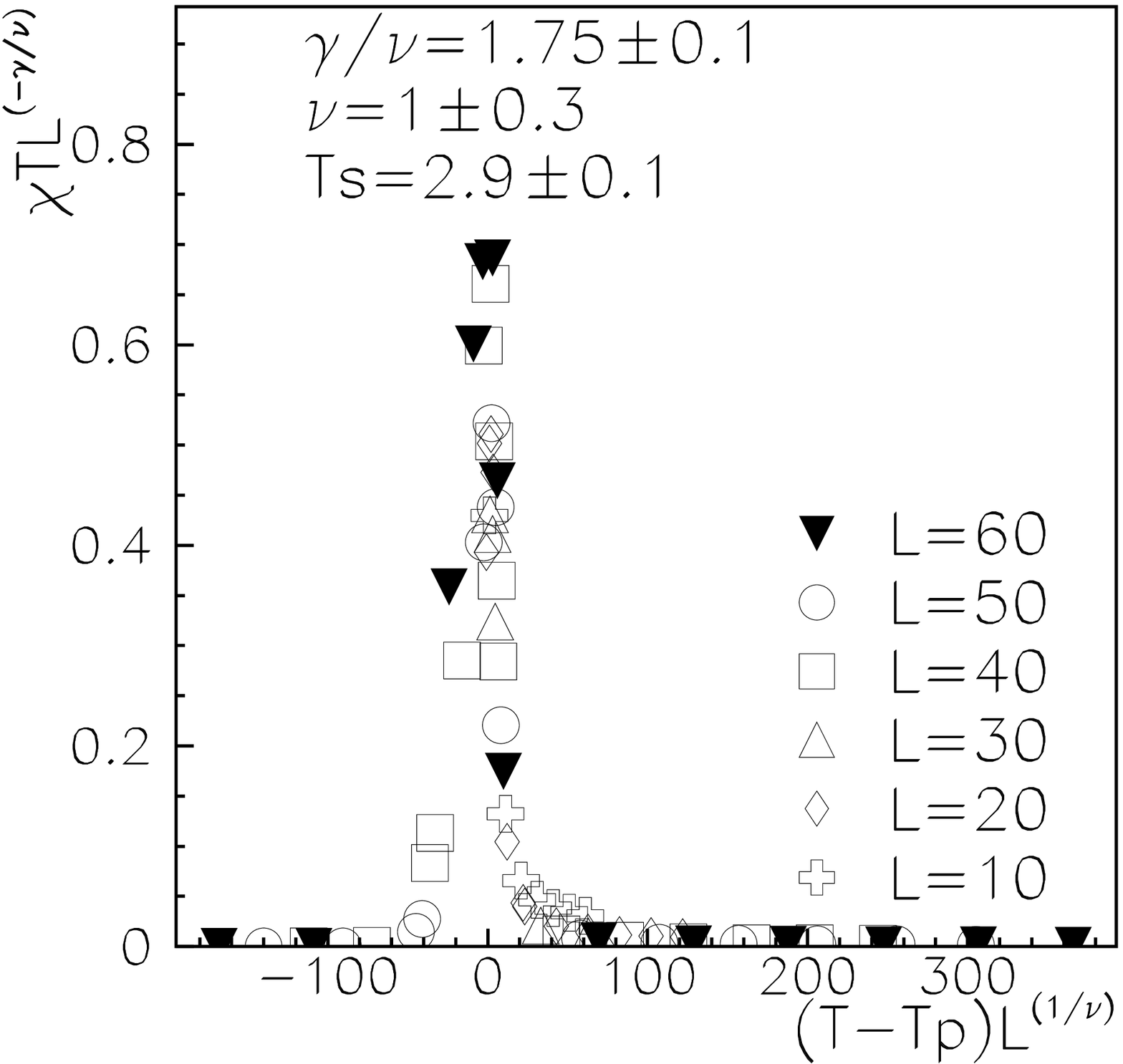 } }
\caption{PSG model with $s=2$: Data collapse for $\chi$ for systems sizes 
$L=10\div 60$. Temperatures ate in $J/k_B$ units. The scaling parameters are given in the 
figure.}
\label{sca_s2_1}
\end{center}
\end{figure}

\begin{figure}
\begin{center}
\mbox{ \epsfxsize=8cm \epsffile{ 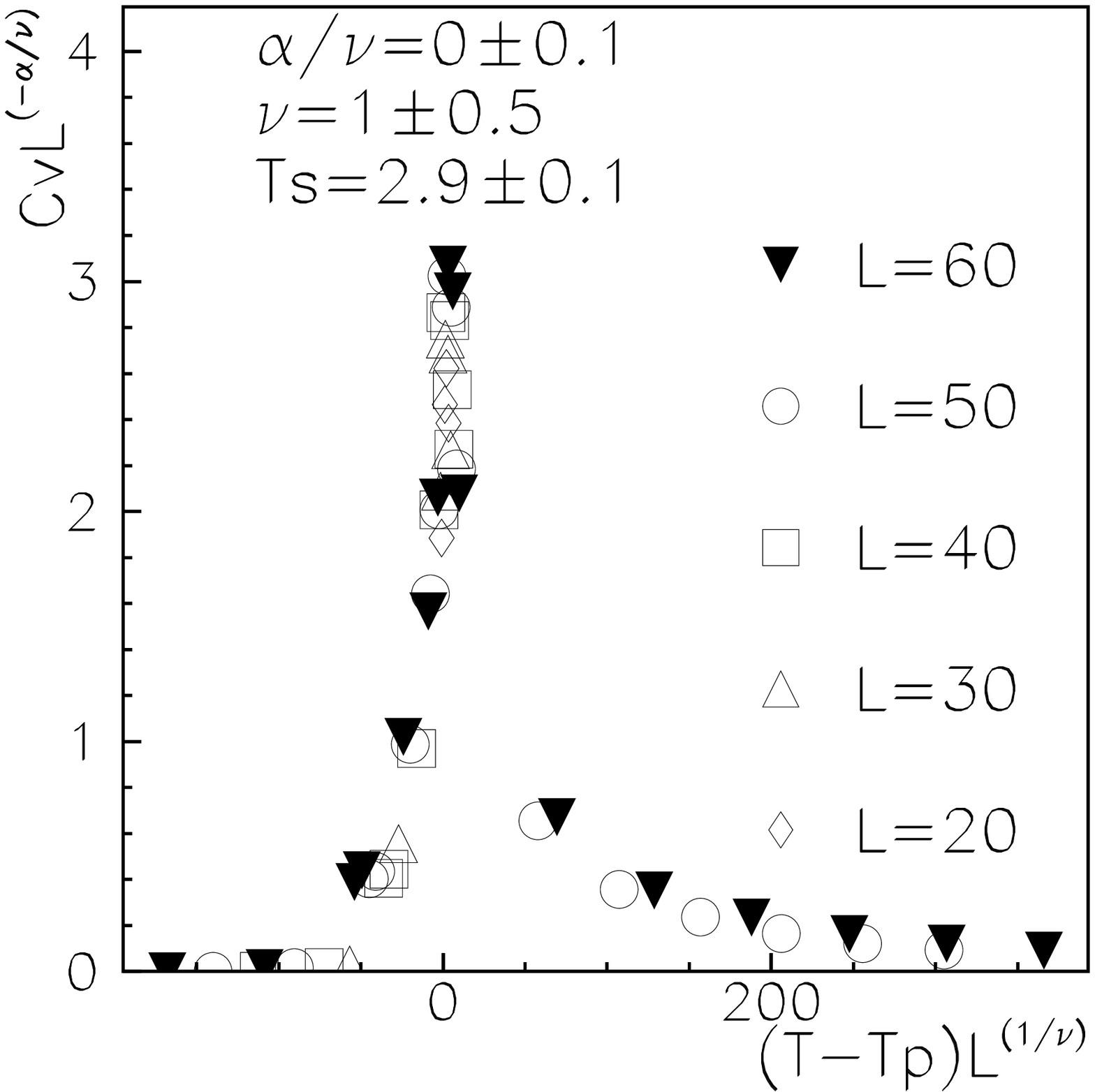 } }
\caption{PSG model with $s=2$: Data collapse for $C_H$ for systems sizes 
$L=10\div 60$. Temperatures ate in $J/k_B$ units. The scaling parameters are given in the 
figure.}
\label{sca_s2_2}
\end{center}
\end{figure}

In Figs. \ref{sca_s2}, \ref{sca_s2_1}, \ref{sca_s2_2} 
we show the data collapses
for system sizes $L=10\div 60$.
The estimated scaling parameters are \cite{errors}
given in Tab. \ref{thermo2}

The estimated critical exponents are compatible, within the errors, with the expected
values for a Potts model with $s=2$ state, {\it i.e.} an Ising model,
in 2D:
$\alpha=0$, $\beta=1/8=0.125$, $\gamma=7/4=1.75$ and $\nu=1$ \cite{Onsager}.
The estimated critical temperature is $k_BT_s/J=2.95\pm 0.15$.

\begin{figure}
\begin{center}
\mbox{ \epsfxsize=8cm \epsffile{ 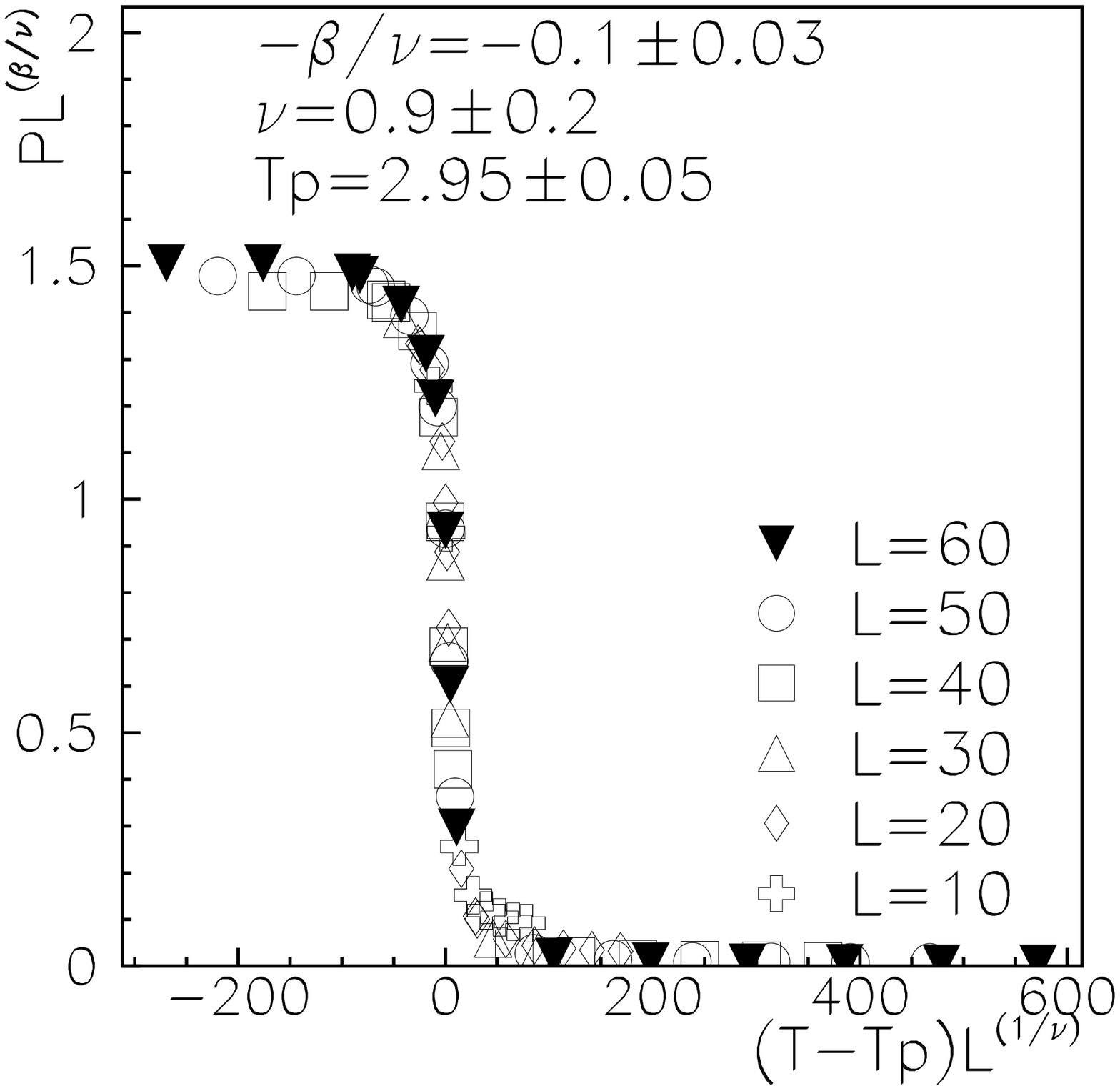 } }
\caption{
PSG model with $s=2$: Data collapses for $P$ for systems sizes $L=10\div 60$.
Temperatures are in $J/k_B$ units. The scaling parameters are given in the 
figure. In figure indexes $p$ are omitted for the critical exponents. 
}
\label{sca_P}
\end{center}
\end{figure}

\begin{figure}
\begin{center}
\mbox{ \epsfxsize=8cm \epsffile{ 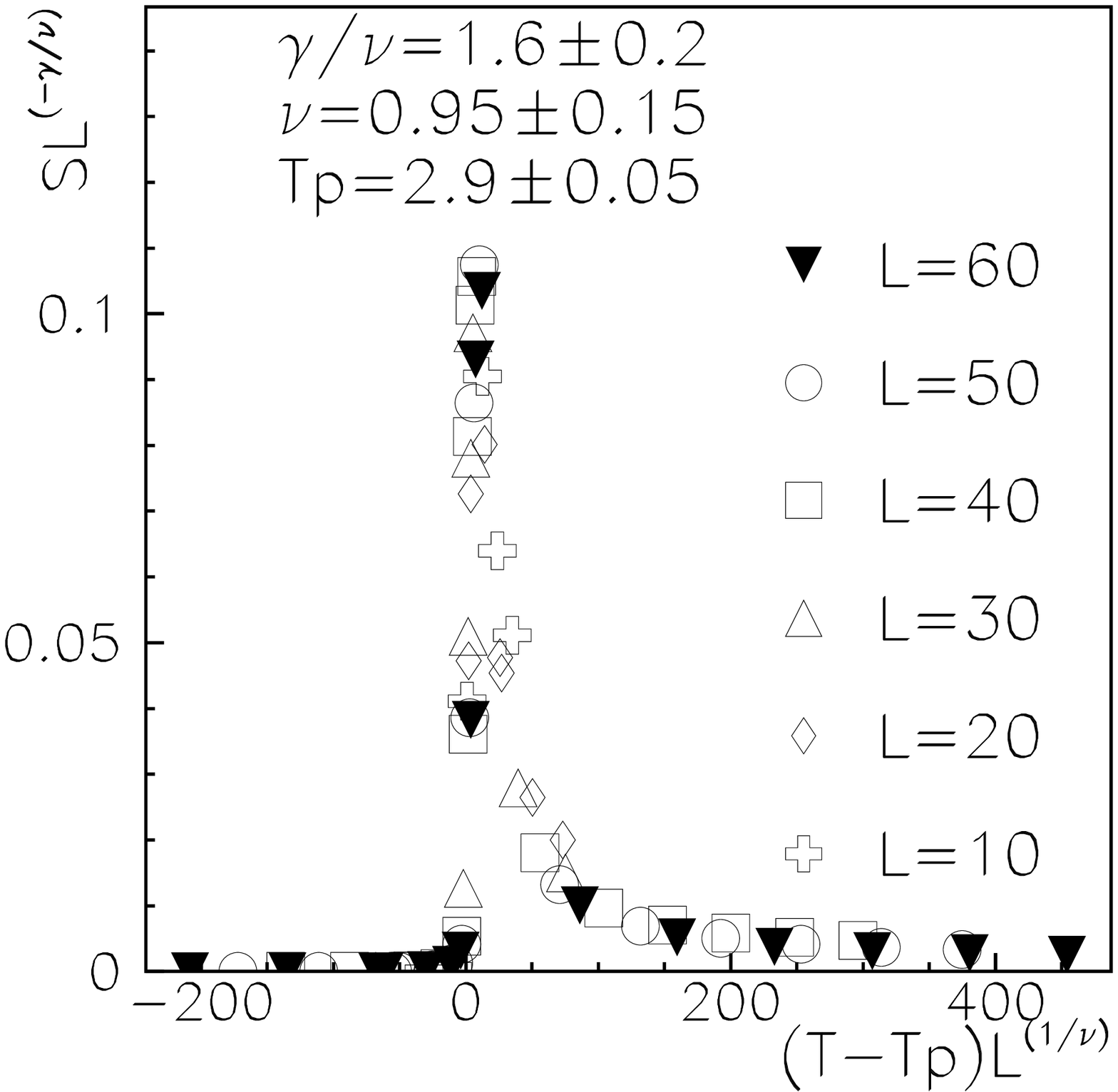 } }
\caption{
PSG model with $s=2$: Data collapses for $S$ for systems sizes $L=10\div 60$.
Temperatures are in $J/k_B$ units. The scaling parameters are given in the 
figure. In figure indexes $p$ are omitted for the critical exponents. 
}
\label{sca_S}
\end{center}
\end{figure}

For the FK-CK percolation quantities we  have described the critical behavior
of $P$ and $S$ introducing a percolation set of critical exponents
($\alpha_p$, $\beta_p$, $\gamma_p$ and $\nu_p$) defined by the relations
\begin{equation}
\xi_p\sim|T-T_p|^{-\nu_p}
\label{nu_p}
\end{equation}
where $\xi_p$ is the connectedness length of the clusters and $T_p=
\lim_{L\rightarrow \infty} T_p(L)$ with $T_p(L)$ finite size 
percolation temperature,
\begin{equation}
P\sim |T-T_p|^{\beta_p}\sim \xi_p^{-\beta_p/\nu_p} ~ ,
\label{beta_p1}
\end{equation}
\begin{equation}
S\sim|T-T_p|^{-\gamma_p}\sim \xi_p^{\gamma_p/\nu_p} ~ .
\label{gamma_p}
\end{equation}
Standard scaling analysis for finite systems \cite{Stauffer_Aha}
is applied also in this
case and the results are summarized in Figs. \ref{sca_P}, \ref{sca_S}.
The estimated scaling parameters are given in Tab. \ref{perco2}.

All the estimated exponents are compatible, within the errors, to the
corresponding thermodynamical parameters for the 2D Ising model and
the numerical estimate for the percolation temperature is $k_B T_p/J=
2.925 \pm 0.075$ consistent with the estimates of $T_s$.

\subsection{Results for $s=7$ and $s=50$}

\begin{figure}
\begin{center}
\mbox{ \epsfxsize=8cm \epsffile{ 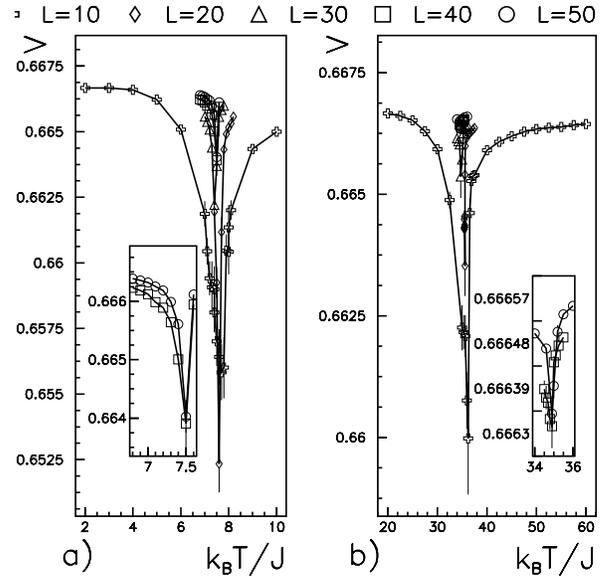 } }
\caption{ PSG model: Binder parameter $V$ vs. dimensionless temperature
$k_B T/J$ for (a) $s=7$ and (b) $s=50$, for $L=10\div 50$. Insets show the
particulars for $L=40$, 50.
Where not shown, the errors are smaller than the symbol sizes.
Lines are only guides for the eyes.
}
\label{V14_100}
\end{center}
\end{figure}

In Fig. \ref{V14_100} we show the Binder parameter $V$ for the PSG
model with $s=7$ and $s=50$ for system sizes $L=10\div 50$.
The fact that $V$ has a non vanishing minimum for every size revels that there
is a first order phase transition. In this case there is no diverging length,
therefore the scaling analysis cannot be applied. This kind of transition is
characterized by the finite size relations \cite{Bin88}
\begin{equation}
C_H(T_s(L),L)\simeq \mbox{max}_T(C_H(T,L))\sim L^D
\label{max_Cv}
\end{equation}
(where $D$ is the Euclidean dimension)
for the maximum of finite size $C_H(L)$ (see Tab. \ref{maxCh}) 
and the relation
\begin{equation}
T_{\mbox{max}}(L)-T_{\mbox{max}}(\infty)\sim L^{-D}
\label{Tp(L)}
\end{equation}
where $T_{\mbox{max}}(L)$ is the temperature of the maximum
of $C_H$ (or of the mean cluster size $S$) for the size $L$ and
$T_{\mbox{max}}(\infty)$ is the corresponding value in the
thermodynamical limit, {\it i.e.} the corresponding
transition temperature $T_s(s)$ (or $T_p(s)$).
Therefore $T_s(s)$ and $T_p(s)$ can be evaluated by linear fits with
one free parameter. The data are given in Tab. \ref{max} and the 
results are for $s=7$ $T_s=T_p=7.5\pm 0.1$ and for $s=50$ 
$T_s=T_p=35.0\pm 0.1$.

\begin{figure}
\begin{center}
\mbox{ \epsfxsize=8cm \epsffile{ 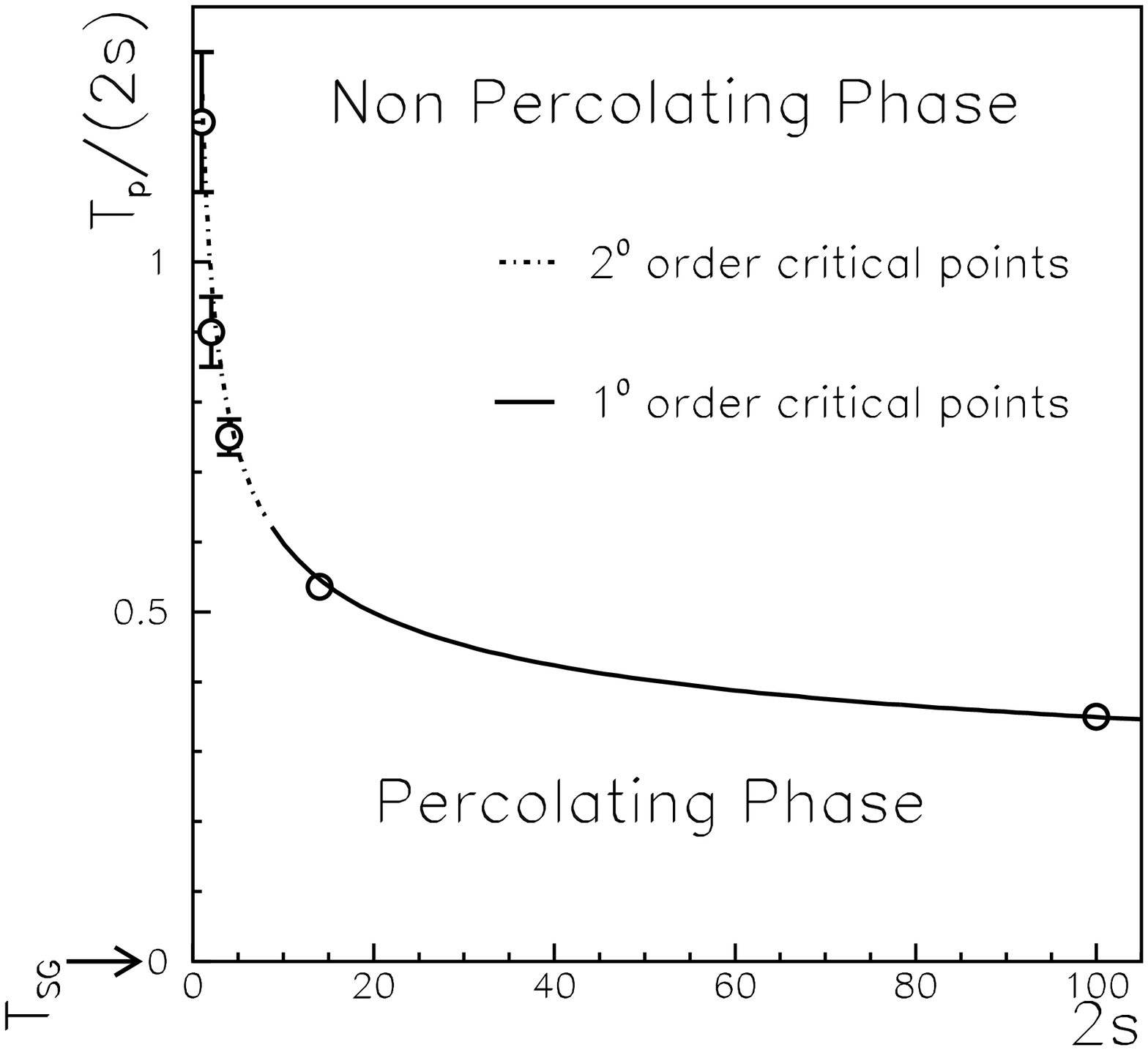 } }
\caption{PSG model: Numerical phase diagram in 2D. The data are fitted with
$T_p/(2s)=a/(\ln(1+\sqrt{2sa}))$ with $a=0.803 \pm 0.003$ (choosing $J=k_B$).
Data for $s=1/2$ and $s=1$ are from Ref.[25]. 
Where not shown, the errors are smaller than the symbol sizes.
}
\label{phase_d}
\end{center}
\end{figure}

The results are summarized in the phase diagram in
Fig.\ref{phase_d}: For every $s$ the high temperature phase is disordered and
non percolating; decreasing the temperature there is a second or first order
phase transition (depending on $s$) at $T_p$ corresponding to the percolation
of FK-CK
clusters and to the ordering of Potts variables; at lower temperature there is
the SG transition (that in 2D occurs at $T=0$\cite{BathY}).
For a fixed realization of $\{\epsilon_{ij}\}$ it is possible to
show \cite{Cataudella_et_al} that each critical point is
characterized by a diverging critical length. At $T_p(s)$
diverges the linear size of FK-CK clusters associated with the pair connectedness
function, while at $T_{SG}(s)$ diverges the linear size of correlated regions.

It is interesting to note that the behavior of $T_p(s)$ can be obtained
from the exact expression of the
transition temperature of a ferromagnetic $2s$-state Potts model \cite{Wu}
only by renormalizing the number of states, {\it i.e.}
\begin{equation}
\frac{T_p}{2sa}=\frac{1}{\ln(1+\sqrt{2sa})}
\end{equation}
with $a=0.803 \pm 0.003$ (choosing $J=k_B$).

\section{Comparison with spin flip dynamics}

The MC dynamics used to study the equilibrium properties of the PSG
model is the Swendsen and Wang (SW) cluster MC dynamics.
The SW dynamics is performed in 
two steps. The first steps is to construct the FK-CK cluster configuration $C$, 
given an Ising and a Potts spin configuration $\{S_i,\sigma_i\}$, activating 
bonds with the probability in eq. (\ref{p}) between NN sites when both 
Ising and Potts spins satisfy the interaction and with zero probability 
otherwise. The second step consists in reversing all the spins in a cluster at 
the same time with probability $1/2$, for each cluster. The sequence of the first
and of the second steps applied to the whole system constitutes a MC step, that is the
chosen unit of time. 

This dynamic completely overcomes the problem of critical slowing down for the
unfrustrated spin models \cite{SW}, while turns out to suffer of diverging
correlation time if applied to frustrated systems near a critical point.
This inefficiency is a consequence of the fact that the FK-CK clusters used in
the SW dynamics do not represent anymore in frustrated models the regions of
correlated spins near a critical point \cite{CdLMP} and their percolation temperature
$T_p$ is greater than the critical temperature. In particular, this is true for
the SG model for which an efficient cluster MC dynamics does not yet exist except 
for 2D \cite{SW2D}, while efficient cluster dynamics have been proposed 
for systems with frustration but without disorder \cite{PRL_PRE,ff3d}.

Nevertheless in SG for temperatures well above the critical temperature $T_{SG}$
and near $T_p$ the SW dynamics is still efficient, consistently with the general 
observation that the cluster dynamics are efficient at least for temperatures
above the percolation temperature \cite{ff3d}.

On the other hand in Ref.\cite{breve} we have shown that the local spin-flip (SF) MC 
dynamics \cite{Bin88} in the case of PSG model with $s=2$ is characterized near $T_p$ 
by diverging correlation times. 
To compare the efficiency of SW dynamics to that of SF dynamics 
we have studied for the case $s=2$
the correlation functions at the equilibrium, that for a generic  observables
$A$ is defined as
\begin{equation}
f_A(t)=\left[\frac{\langle \delta(t+t_0) \delta(t_0) \rangle}{\langle
\delta(t_0)^2 \rangle }\right] ,
\label{corrM}
\end{equation}
where $\delta(t)=A(t)-\langle A\rangle$ and $t_0$ is the equilibration time.
As observables we have choose the  Potts order parameter $M$
and the energy $E$ of the whole system.

We have also studied the time dependent nonlinear susceptibility
for a quenched interaction configuration
\begin{equation}
\chi_{SG}(t)=\frac{1}{N} \left\langle \left[\sum_i
S_i(t+t_0) S_i(t_0)\right]^2\right\rangle
\label{chi_sg_t}
\end{equation}
(where $N$ is the total number of spins).
The normalized correlation function is
\begin{equation}
f_\chi=\frac{\chi_{SG}(t)-\chi_{SG}(t=\infty)}{\chi_{SG}(0)
-\chi_{SG}(t=\infty)}
\end{equation}
with $\chi_{SG}(0)=N$.

For both the SW and the SF dynamics we 
have measured the integral correlation time defined as
\begin{equation}
\tau_{int}=
\lim_{t_{max}\rightarrow \infty}\frac{1}{2}+\sum_{t=0}^{t_{max}}f(t)
\label{int}
\end{equation}
where $f$ is the generic correlation function. We have considered 
systems with lattice sizes $L\leq 30$ at temperatures above
and below $T_p$. 
The data for the SF dynamics are averaged over 32 different
quenched interaction configurations, since the local updating of this
dynamics strongly depends on the local fluctuation of the frustration.
On the other hand the results on global SW dynamics turns out to be
``robust" respect the interaction configuration average, in the sense 
that the fluctuations of $\tau_{int}$ are within the errors estimated 
on the basis of a single interaction configuration analysis.

The simulations have been done
with an annealing method, {\it i.e.} with a slow cooling of the
system at each temperature.
For the SW cluster dynamics $5~10^3$ MC steps turn out to be 
enough to equilibrate the system at the 
considered temperatures and the averages are done using the 
the data for the following $5~10^4$ MC steps.
For the SF dynamics we have discarded the first $10^4$ MC steps 
(defined as the local update of any spin in the system) and 
recorded the data for $5~10^5$ MC steps.

In Tab. \ref{tau} we show the results for $L=30$. 
Analogous results have been found for smaller systems.

The data show that, while the SF correlation times for $M$ and $E$ 
grow abruptly near $T_p$ where both thermodynamical and percolation
transition occur, the SW correlation times only show a slow trend to increase
for decreasing temperatures, being smaller than the correspondent 
SF data at least of an order of magnitude.
Even for $\tau_{\chi}$, that for SF starts to be non-zero below $T_p$,
the SW dynamics shows smaller correlation times.
For temperatures well below $T_p$ it is possible to see that the SW dynamics
is characterized by long autocorrelation times, as the SF dynamics. 
Therefore, at least for temperatures not too much  below $T_p$, the SW dynamics  
turns out to be more efficient than local SF dynamics. In particular, near the
thermodynamical transition at $T_p$ the SW dynamics completely overcomes the 
critical slowing down problem in the PSG model, as for the unfrustrated
models, even if frustration is present via the random interactions of the
Ising spins. 

\section{Conclusions}

Precursor phenomena characterize the paramagnetic phase of spin glass. In particular,
experiments and local Monte Carlo simulations show
the presence of stretched exponential autocorrelation functions well above the 
SG transition temperature $T_{SG}$ \cite{Campbell}.
The relation of the onset $T^*$ of these precursor phenomena to any
thermodynamical transition and the localization of $T^*$ are still matters of 
debate \cite{Campbell,breve}.
Many numerical evidences on disordered \cite{Campbell} and deterministic 
frustrated 
models \cite{Fierro,FF}
have shown that $T^*$ is consistent with the percolation temperature $T_p$ of 
the the
Kasteleyn - Fortuin and Coniglio - Klein clusters.
In particular, a generalization of the $\pm J$ Ising SG  to a $2s$-state Potts 
SG,
that recovers the SG for $s=1$, has shown that for $s=2$, as for $s=1$ 
\cite{Campbell},
the $T^*=T_p$ hypothesis is numerically verified \cite{breve}.
In this paper, using very efficient cluster Monte Carlo dynamics,  
we have shown that in 2D for any $s\neq 1$ the percolation
transition corresponds to a real thermodynamical transition in the 
universality class
of the $s$-state ferromagnetic Potts model.
In particular, we have considered the case $s=2$ and the cases $s=7$, 50 where 
at $T_p(s)$ a second order and, respectively, a first order phase transition 
occurs.
Exact renormalization group calculations on
hierarchical lattices \cite{Pezzella} and mean field analysis for a version of 
the model
without disorder \cite{dLP} have shown the same scenario.

All these results suggest that the percolation transition may play a role
in the precursor phenomena even in the SG case ($s=1$), where no thermodynamical
transition occurs at $T_p$. This idea arises from the observation that, even 
in SG, 
below $T_p$ the frustration starts to be manifested on all the length scales 
by means
of the FK-CK clusters, that cannot include frustrated loops. Therefore the 
scenario
presented is that for $s\neq 1$ at the percolation temperature $T_p$ there is a 
thermodynamical transition with
associated dynamical anomalies that "vanishes" for $s=1$ leaving the dynamical
behavior unchanged.


\begin{table}
\caption{Estimated critical exponents and critical temperatures $T_s$ 
for thermodynamical quantities $M$, $\chi$ and $C_H$ for the PSG model with $s=2$.
}
\begin{tabular}{l|c c c c c }
   ~   & $\nu$        & $\alpha/\nu$ & $\beta/\nu$     & $\gamma/\nu$   & $k_BT_s/J$ \\
\hline
$M$    & $0.9\pm 0.3$ & ~            & $0.13\pm 0.06$ & ~              & $3.0\pm 0.1$\\
$\chi$ & $1.0\pm 0.3$ & ~            & ~              & $1.75\pm 0.10$ & $2.9\pm 0.1$ \\
$C_H$  & $1.0\pm 0.5$ & $0.0\pm 0.1$ & ~              & ~              & $2.9\pm 0.1$
\label{thermo2}
\end{tabular}
\end{table}

\begin{table}
\caption{Estimated percolation exponents and percolation temperature $T_p$ 
for $P$ and $S$ for the PSG model with $s=2$.
}
\begin{tabular}{l|c c c c }
   ~   & $\nu_p$          & $\beta_p/\nu_p$ & $\gamma_p/\nu_p$ & $k_BT_p/J$ \\
\hline
$P$    & $0.9\pm 0.2$      & $0.10\pm 0.03$  & ~                & $2.95 \pm 0.05$\\
$S$    & $0.95\pm 0.15$   & ~               & $1.6\pm 0.2$     & $2.90\pm 0.05$ 
\label{perco2}
\end{tabular}
\end{table}

\begin{table}
\caption{Maxima of $C_H$ for $s=7$ and $s=50$ for $L=30$, 40, 50.}
\begin{tabular}{l|c c c }
$L$            & 30 & 40 & 50 \\
\hline
$\mbox{max}_T C_H(s=7)/L^2$  & $27\pm 3$    & $26\pm 3$   & $33\pm 3$ \\
$\mbox{max}_T C_H(s=50)/L^2$ & $7.4\pm 0.6$ & $8.4\pm 0.9$& $7.3\pm 0.7$
\label{maxCh}
\end{tabular}
\end{table}

\begin{table}
\caption{Temperatures (in $J/k_B$ units) of maxima of $C_H$ and $S$ for $s=7$ and 
$s=50$ for $L=10$, 20, 30, 40, 50.}
\begin{tabular}{l|c c c c c }
$L$     & 10 & 20 & 30 & 40 & 50  \\
\hline
$T_{\mbox{max }C_H}(s=7)$  & 7.8 & 7.6 & 7.5 & 7.5 & 7.5  \\
$T_{\mbox{max }S}(s=7)$    & 7.9 & 7.7 & 7.5 & 7.5 & 7.52 \\
$T_{\mbox{max }C_H}(s=50)$ & 36.56& 35.57& 34.92& 34.90& 35.00 \\
$T_{\mbox{max }S}(s=50)$   & 36.56& 35.60& 35.00& 35.00& 35.00 
\label{max}
\end{tabular}
\end{table}

\begin{table}
\caption{
Integral correlation times for a PSG model with $s=2$ and $L=30$ for Swendsen and Wang
(SW) cluster MC dynamics and local spin-flip (SF) MC dynamics for temperatures
above and below $T_p(L=30)\simeq 2.95 J/k_B$.
}
\begin{tabular}{l|c c c }
$k_B T/J$         & 2.75           & 3.00           & 3.25 \\
\hline
$\tau_{M}(SW)$    & $3.08\pm 0.02$ & $3.65\pm 0.03$ & $1.81\pm 0.01$ \\
$\tau_{M}(SF)$    & $76.8\pm 0.3$  & $575.8\pm 0.6$ & $180.3\pm 0.2$ \\
\hline
$\tau_{E}(SW)$    & $9.32\pm 0.02$ & $8.53\pm 0.07$ & $4.45\pm 0.04$ \\   
$\tau_{E}(SF)$    & $19.0\pm 0.1$  & $67.5\pm 0.1$  & $20.9\pm 0.5$  \\
\hline
$\tau_{\chi}(SW)$ & $9.93\pm 0.07$ & $2.423\pm 0.002$ & $1.585\pm 0.001$ \\
$\tau_{\chi}(SF)$ & $16.89\pm 0.02$ & $7.06\pm 0.03$  & $3.97\pm 0.01$
\label{tau}
\end{tabular}
\end{table}

\begin{references}
\bibitem{errors}
The errors are overestimated since their evaluation is done changing a 
parameter and taking the other fixed at the best value.
\end{references}
\end{document}